\documentstyle[11pt]{article}

\skip\footins 5mm plus 2mm minus 2mm
\begin{document}
\setlength{\parindent}{0 mm}

\vspace*{-0.5in}

\vspace*{-0.5in}

\def\bd{\begin{displaymath}}
\def\ed{\end{displaymath}}
\def\be{\begin{equation}}
\def\ee{\end{equation}}
\def\p{\partial}

\newtheorem{theorem}{Theorem}
\newtheorem{conjecture}{Conjecture}
\newtheorem{lemma}{Lemma}
 \newtheorem{proposition}{Proposition}
\newcommand {\ds} {\displaystyle \sum}
\vspace*{-0.5in}

\renewcommand{\thesection}{\arabic{section}}

\large
 \vskip 1 cm \vskip 1cm \vskip 1cm

\centerline { {\bf  THE TODA LATTICE IS SUPER--INTEGRABLE }  }

\

 \vskip 1  cm
\centerline { Maria Agrotis, Pantelis A. Damianou, Christodoulos
Sophocleous}

\vskip .5 cm

 \centerline { Department of
Mathematics  and Statistics} \centerline  { University of Cyprus}

\centerline { P. O. Box 20537, 1678 Nicosia, Cyprus }
\centerline{Email: agrotis@ucy.ac.cy, damianou@ucy.ac.cy,
christod@ucy.ac.cy}
 \vskip 2 cm \centerline { \bf {ABSTRACT }}
\bigskip
{\it  We prove  that the classical, non--periodic Toda lattice  is
super--integrable. In other words, we show that it possesses
$2N-1$ independent constants of motion, where $N$ is the number of
degrees of freedom. The main ingredient of the proof is the use of
some special action--angle coordinates introduced by Moser to
solve the equations of motion. }

\vskip .5 cm


\vskip .5 cm


\section{Introduction}
The Toda lattice is arguably the most fundamental and basic of all
finite dimensional Hamiltonian integrable systems. It has various
intriguing connections with other parts of mathematics and
physics.

 The Hamiltonian of the Toda lattice is given by  \be
  H(q_1, \dots, q_N, \,  p_1, \dots, p_N) = \sum_{i=1}^N \,  { 1 \over 2} \, p_i^2 +
\sum _{i=1}^{N-1} \,  e^{ q_i-q_{i+1}}  \ . \label{a1} \ee

This type of Hamiltonian was considered first by Morikazu Toda
\cite{toda}.  Equation (\ref{a1}) is known as the classical,
finite, non--periodic Toda lattice to distinguish the system from
the many and  various other versions, e.g.,  the relativistic,
quantum, infinite,  periodic etc.  The integrability of the system
was established in 1974 independently by Flaschka
\cite{flaschka1}, H\'enon  \cite{henon} and Manakov
\cite{manakov}.
 The original Toda
lattice can be viewed as a discrete version of the Korteweg--de
Vries equation. It is called a lattice as in atomic lattice since
interatomic interaction was studied. This  system also appears  in
Cosmology. It appears also in the work of Seiberg and Witten on
supersymmetric Yang--Mills theories and it has applications in
analog computing and numerical computation of eigenvalues. But the
Toda lattice is mainly a theoretical mathematical model which is
important due to the rich mathematical structure encoded in it.

Hamilton's equations become

\bd
\begin{array}{lcl}
\dot q_j =p_j    \\
\dot p_j=e^{ q_{j-1}-q_j }- e^{q_j- q_{j+1}}  \ .
\end{array}
\ed

\smallskip
\noindent The system is  integrable. One can find a set of
independent functions $\{  H_1, \dots,  H_N \} $  which are
constants of motion for Hamilton's equations. To  determine the
constants of motion, one uses Flaschka's transformation: \be
  a_i  = {1 \over 2} e^{ {1 \over 2} (q_i - q_{i+1} ) }  \ \ , \ \ \ \ \ \ \ \ \ \ \ \ \ \
             b_i  = -{ 1 \over 2} p_i  \ .     \label{a2}
\ee

\smallskip
\noindent Then

\be
\begin{array}{lcl}
 \dot a _i& = & a_i \,  (b_{i+1} -b_i )    \\
   \dot b _i &= & 2 \, ( a_i^2 - a_{i-1}^2 ) \ .  \label{a3}
\end{array}
\ee

\smallskip
\noindent These equations can be written as a Lax pair  $\dot L =
[B, L] $, where $L$ is the Jacobi matrix

\bd
 L= \pmatrix { b_1 &  a_1 & 0 & \cdots & \cdots & 0 \cr
                   a_1 & b_2 & a_2 & \cdots &    & \vdots \cr
                   0 & a_2 & b_3 & \ddots &  &  \cr
                   \vdots & & \ddots & \ddots & & \vdots \cr
                   \vdots & & & \ddots & \ddots & a_{N-1} \cr
                   0 & \cdots & & \cdots & a_{N-1} & b_N   \cr } \ ,
\ed

\smallskip
\noindent and

\bd
    B =  \pmatrix { 0 & a_1 & 0 & \cdots & \cdots &  0 \cr
                 -a_1 & 0 & a_2 & \cdots & & \vdots  \cr
                    0  & -a_2 & 0 & \ddots &  & \cr
                    \vdots &  & \ddots & \ddots & \ddots & \vdots \cr
                     \vdots & & &  \ddots & \ddots & a_{N-1} \cr
                     0 & \cdots &\cdots &  & -a_{N-1}  & 0 \cr } \ .
\ed

\bigskip
\noindent This is an example of an isospectral deformation; the
entries of $L$ vary over time but the eigenvalues  remain
constant.
 It follows that the  functions $ H_j={1 \over j} {\rm tr} \, L^j$ are  constants of
motion. The mappings (\ref{a2}) is a projection from ${\bf
R}^{2N}$ to ${\bf R}^{2N-1}$ and it is clearly not one--to--one.
Replacing the vector ${\bf q}$ by ${\bf q}+c$ gives the same
image.

The supper--integrability of this type of  systems should be
expected due to their  dispersive  asymptotic behavior. However,
the construction of integrals is not typically a trivial task. In
the case of the open Toda lattice, asymptotically the particles
become free as time goes to infinity with asymptotic momenta being
the eigenvalues  of the Lax matrix. Therefore, the system behaves
asympotically  like a system of free particles which is
super--integrable.

The super--integrability of the Toda lattice for $N=2$ was
established  in \cite{damianou6}.  The additional integral in
\cite{damianou6} was obtained using Noether's theorem. We give the
formula for the additional integral.

In the case of two degrees of freedom the potential is simply \bd
V(q_1, q_2)=e^{q_1-q_2} \ , \ed and the procedure of Noether
produces the following three integrals:

\bd H_1=-{1 \over 2} (p_1+p_2)  \ , \qquad J_1=(p_1-p_2)^2+4
e^{q_1-q_2} \ , \ed

\be I_1={ p_1-p_2 + \sqrt{ J_1} \over p_1-p_2-\sqrt{ J_1}} \exp
\left( \sqrt{J_1} {q_1+q_2 \over p_1+p_2} \right) \ . \label{a0}
\ee Note that $H=H_1^2+ {1\over 4} J_1 $ and
 that the function $ G={ q_1 +q_2 \over p_1+p_2}$ which appears in the exponent of
 $I_1$ is a time function, i.e., it  satisfies $\{G,H \}=1$. We
 note that the integral $I_1$  remains an integral if we add a
 real constant to $q_1+q_2$, an observation that will be important
 later on.

The existence of the integral $I_1$  shows that the two degrees of
freedom Toda lattice is super--integrable with three integrals of
motion $\{H_1, J_1, I_1 \}$. As we will see, the complicated
integral $I_1$ has a simple expression  if one uses Moser's
coordinates.

\section{Moser's solution of the Toda lattice}

Moser's beautiful  solution  of the open Toda lattice uses the
Weyl function $f(\lambda)$ and an old (19th century) method of
Stieltjes which connects the continued fraction  of $f(\lambda)$
with its partial fraction expansion.
 The key ingredient is the
map which takes the $(a,b)$ phase space of tridiagonal Jacobi
matrices to a new space of variables $(\lambda_i, r_i)$
 where $\lambda_i$ is an eigenvalue of the Jacobi matrix and $r_i$ is related to the residue of some  rational functions that appear
in the solution of the equations.
 We present a brief outline of Moser's construction.

Moser in \cite{moser} introduced the resolvent \bd
R(\lambda)=\left( \lambda I - L \right)^{-1} \ , \ed

 and defined the Weyl function
\bd f(\lambda)=\left( R(\lambda)e_N, \  e_N \right) \ , \ed

where $e_N=(0,0, \dots, 0, 1)$.

The function $f(\lambda)$ has a simple pole at $\lambda=
\lambda_i$. For the purpose of this paper, and this is also
observed by Moser, we define $f$ by the formula
\begin{equation} \label{pf}
f(\lambda)=\left( \sum_{i=1}^{N} \frac{r_i^2}{\lambda-\lambda_i}
\right)/(\sum_{i=1}^N r_i^2) \ .
\end{equation}

 The differential equations in the variables $(\lambda, r)$ take
a particularly simple form:

\begin{eqnarray}
\dot{\lambda}_i& = & 0 \nonumber\\
\dot{r}_i & = & -\lambda_i \,  r_i\;. \label{a7}
\end{eqnarray}

These equations show that  $(\lambda_i, \, \log{r_i} )$ are
action--angle variables for the Toda lattice.

 The variables $a_i^2$, $b_i$ may be expressed as rational
functions of $\lambda_i$ and $r_i$ using a continued fraction
expansion of $f(\lambda)$ which dates back to Stieltjes. Since the
computation of the continued fraction from the partial fraction
expansion is a rational process the solution is expressed as a
rational function of the variables $(\lambda_i, \ r_i)$. The
procedure is as follows:

The $R_{NN}$ element of the resolvent, as defined previously,
takes the following continued fraction representation:

\begin{equation}
f(\lambda)=\frac{1}{\lambda-b_N-\frac{a_{N-1}^2}{\lambda-b_{N-1}-\frac{a_{N-2}^2}{\frac{\vdots}
{\lambda-b_2-\frac{a_1^2}{\lambda-b_1}}}}}   \ .
\end{equation}

The function $f(\lambda)$ has $N$ simple poles at the eigenvalues
of the Lax pair matrix $L$. Therefore, its partial fraction
expansion has the form:

\begin{equation} \label{pf}
f(\lambda)=\left( \sum_{i=1}^{N} \frac{r_i^2}{\lambda-\lambda_i}
\right)/(\sum_{i=1}^N r_i^2) \ ,
\end{equation}
where the residue of $f(\lambda)$ at $\lambda=\lambda_i$ is
$r_i^2/(\sum_{i=1}^N r_i^2)$. Stieltjes described a procedure that
allows one to  express $a_i$ and $b_i$ in terms of $\lambda_1,
\ldots,\lambda_N$ and $r_1,\ldots,r_N$. We briefly describe the
method. We expand the partial fraction expansion of $f(\lambda)$
as given in (\ref{pf}) in a series of powers of
$\frac{1}{\lambda}$. We obtain,

\bd f(\lambda)=\left[ \sum_{j=0}^{\infty} \frac{ \sum_{i=1}^{N}
r_i^2 \lambda_i^j}{\lambda^{j+1}} \right]/(\sum_{i=1}^N r_i^2)\ .
\ed The coefficient  of $\lambda^{j+1}$ is denoted by $c_j$ and
equals,

\bd c_j=\left( \sum_{i=1}^{N} r_i^2 \lambda_i^j \right)
/(\sum_{i=1}^N r_i^2), \quad j=0, 1, \ldots. \ed The formulas of
Stieltjes involve certain $i \times i$ determinants which we now
define,

\bd
A_i=\left| \begin{array}{cccc} c_0 & c_1 & \ldots & c_{i-1} \\
                    c_1 & c_2 & \ldots & c_{i} \\
                    \vdots & & &\\
                    c_{i-1} & c_{i} & \ldots & c_{2i-2}
    \end{array} \right|, \qquad B_i=\left| \begin{array}{cccc} c_1 & c_2 & \ldots & c_{i} \\
                    c_2 & c_3 & \ldots & c_{i+1} \\
                    \vdots \\
                    c_{i} & c_{i+1} & \ldots & c_{2i-1}
    \end{array} \right|.
\ed The formulas that give the relation between the variables
$(a,b)$ and $(r,\lambda)$ are,

\begin{eqnarray*}
&& a_{N-i}^2=\frac{A_{i-1} A_{i+1}}{A_i^2}, \hspace{2.32cm}
i=1,\ldots,N-1
\\
&&  b_{N+1-i}=\frac{A_i
B_{i-2}}{A_{i-1}B_{i-1}}+\frac{A_{i-1}B_{i}}{A_i B_{i-1}}, \quad
i=1,\ldots,N
\end{eqnarray*}
where $A_0=1, B_0=1, B_{-1}=0.$

For example, in the case  $N=2$

\bd A_1=c_0, \quad A_2=c_0 c_2-c_1^2, \quad B_1=c_1, \quad B_2=c_1
c_3-c_2^2 \ed and therefore \bd a_1^2=A_2, \quad
b_1=\frac{A_2}{B_1}+\frac{B_2}{A_2 B_1}, \quad b_2=B_1 \ . \ed
Thus,
\begin{eqnarray}
&& a_1^2=\frac{r_1^2 r_2^2 (\lambda_2-\lambda_1)^2}
{(r_1^2+r_2^2)^2}\nonumber
\\
&& b_1=\frac{r_1^2 \lambda_2+r_2^2 \lambda_1}{r_1^2+r_2^2}  \label{t1}\\
&& b_2=\frac{r_1^2 \lambda_1+r_2^2 \lambda_2}{r_1^2+r_2^2} \
.\nonumber
\end{eqnarray}
One can check that the differential equations $\dot{r_i}=-r_i
\lambda_i$, for $i=1,2$ correspond via transformation (\ref{t1})
to the $A_2$ Toda equations

\begin{eqnarray*}
&& \dot{a_1}=a_1(b_2-b_1)  \\
&& \dot{b_1}=2 a_1^2   \\
&& \dot{b_2}=-2 a_1^2 \ .
\end{eqnarray*}

As Moser notes, it is not too hard to obtain explicit expressions
for $N=3$ but the general case is quite complicated. With the
exception of $N=2$ the $a_i$ are not rational functions of
$(\lambda_i, r_i)$ but the $a_i^2$ are. In general one can express
(at least in theory) the functions $a_i, b_i$ in terms of $r_i,
\lambda_i$. Again, the function is not one--to--one. In fact, if
one replaces the vector ${\bf r}$ with $ \alpha {\bf r}$ the
result is the same.

 Finally, we comment on the Poisson
brackets in the new coordinates.  The multi--hamiltonian structure
of the Toda lattice was developed in \cite{damianou1} and
\cite{damianou2} using master symmetries. The analogous results in
$(\lambda, r)$ coordinates are due to  Feybusovich and Gekhtman
\cite{fayb}.  The Poisson brackets  project onto some rational
brackets in the space of Weyl functions and in particular, the
Lie--Poisson bracket of the Toda lattice  corresponds to the
Atiyah--Hitchin bracket \cite{atiyah}.
 In general, one  constructs a sequence
of Poisson brackets on the space $(\lambda_i, r_i)$ whose image
under the inverse spectral transform corresponds to the standard
Toda hierarchy. A rational function of the form ${q(\lambda) \over
p(\lambda)}$ is determined uniquely by the distinct eigenvalues of
$p(\lambda)$, $\lambda_1, \dots, \lambda_n$ and values of $q$ at
these roots. The residue  is equal to
 ${ q(\lambda_i) \over p^{\prime}(\lambda_i)}$
and therefore we may choose

\bd \lambda_1, \dots, \lambda_n, q(\lambda_1), \dots, q(\lambda_n)
\ed
  as global coordinates on the
space of rational functions (of the form ${q \over p}$ with $p$
having simple roots and $q, p$ coprime). We have to remark that
the image of the Moser map  is a much larger set.

 The  $k$th Poisson bracket is defined by

\bd
\begin{array}{lcl}
\{\lambda_i, \ q(\lambda_i) \}& =&-\lambda_i^k q(\lambda_i)  \nonumber  \\
\{q(\lambda_i),\  q(\lambda_j) \}& = & \{ \lambda_i,\  \lambda_j
\}=0 \ . \nonumber
\end{array}
\ed

The initial Poisson bracket under the Moser map is given
explicitly by

\begin{eqnarray}
\{\lambda_i,\lambda_j\}&=& 0   \nonumber \\
\{r_i,r_j\} &=& 0  \nonumber  \\
\{\lambda_i,  r_j\} &=& \delta_{ij} r_j \;\;\;   i,j=1,\ldots,N \
. \label{a8}
\end{eqnarray}

Similarly, the quadratic Toda bracket,  corresponds to  a bracket
with only non--zero terms $\{ \lambda_i , r_i \}=\lambda_i r_i $.

 The Hamiltonian function in the new coordinates is
$\displaystyle H_2=\frac{1}{2}\sum_{i=1}^{N} \lambda_i^2$. In
other words, taking $H_2$ as the Hamiltonian and using  bracket
(\ref{a8}) gives equations (\ref{a7}).

\section{The Toda lattice is super--integrable}
We now come to the main result of this paper. We define \be
I_j=\left( \frac{r_{j}}{r_{j+1}} \right)^2 e^{F_{j,j+1}}, \
 \ j=1,\ldots,N-1, \label{a9} \ee where

\bd F_{j,j+1}=\frac{2(\lambda_{j}-\lambda_{j+1})}{H_1}
\ln\left(\prod_{i=1}^{N} r_i  \right)\;. \ed It is easily shown,
using equation (\ref{a7})  that $\displaystyle \frac{dI_j}{dt}=0$,
for  $j=1,\ldots, N-1 $ and thus the functions $I_j$  are
constants of motion.

The functions $H_i=\lambda_1^i+ \lambda_2^i+ \dots+\lambda_N^i$
and $I_j, \;i=1, \ldots,N$, $ j=1,\ldots,N-1$ are functionally
independent. In fact, the Jacobian $(2N-1) \times 2N$ matrix of
the functions $H_i$ and $I_j$ has a $(2N-1) \times (2N-1)$
subdeterminant, $d_{N+1}$, which is obtained by deleting the
$(N+1)$-column and is not identically zero. A simple calculation
gives

\bd d_{N+1}=- 2^{N-1} N \frac{r_1^2}{r_{_{N-2}} r_{_{N-1}}
r_{_{N}}^3} \frac{\lambda_1}{H_1} \; e^{F_{1,N}} \prod_{1 \leq i <
j \leq N} (\lambda_i -\lambda_j)\;. \ed Since the eigenvalues of
real Jacobi matrices are distinct, the functions $H_i$ and $I_j$
are independent. We summarize  the results in the following:

 \begin{theorem}
 The Toda lattice with $N$ degrees of freedom posesses $2N-1$ independent
constants of motion, $H_i$, $i=1,\ldots,N$, $I_j$,  $ j=1,
\ldots,N-1$, and is therefore super--integrable.
\end{theorem}

{\it Remark 1}

It is clear that the functions $H_n, \; n=1,\ldots,N$ are in
involution.  Moreover it can be shown that $\{I_i,I_j\}=0,\;
i,j=1,\ldots,N-1$. In addition,  for $ n=1,\ldots,N,
\;\;j=1,\ldots,N-1$

\bd \{ H_n,I_j \}=2c_n \frac{(\lambda_{j} - \lambda_{j+1})}{H_1}
{E}_{j} I_j, \ed where $c_n=1$ for $n=2,\ldots,N$,
$c_1=\frac{N}{N-2}$ and

\bd {E}_{j}=\sum \lambda_{i}^{n-1} - \left( \sum \lambda_i \right)
w(n-2) - 2 \lambda_{j+1} \lambda_{j} \; w(n-3). \ed

 The
sums are taken over all $i$ from $1$ to $N$ where ${i \neq j,j+1}$
The function $w(n)$ symbolizes the full homogeneous polynomial in
$\lambda_j$ and $\lambda_{j+1}$ that have total weight equal to
$n$. For instance, $w(n)=0, n \in - {{\bf Z}^{+}}$, $w(0)=1$,
 $w(1)=\lambda_{j}+ \lambda_{j+1}$,   $w(2)=
\lambda_j^2+\lambda_{j+1}^2+\lambda_j \lambda_{j+1}$, etc.

One can, of course, use the quadratic Toda bracket in $(\lambda_i,
r_i)$ coordinates. We must then take Tr$L=\lambda_1 + \dots +
\lambda_N$ as Hamiltonian. However, in this bracket the $H_i, I_j$
do not form a  finite dimensional algebra.

\bigskip
{\it Remark 2}

We clearly have  $\{H_2,I_j\}=0, \;j=1,\ldots, N-1$,  since $H_2$
is the Hamiltonian  and the functions $I_j$ are constants of
motion.

We define the sets $S_1=\{ H_1,\ldots, H_N \}$ and $S_2=\{
H_2,I_1,\ldots, I_{N-1 } \}$. Then if $f,g \in S_1 \Rightarrow \{
f,g \}=0$ and if $f,g \in S_2 \Rightarrow \{ f,g \}=0$. In other
words the sets $S_1$ and $S_2$ are both maximal sets of integrals
in involution. We therefore have two different sets demonstrating
the complete  integrability of the Toda lattice.

\bigskip
{\it Remark 3}

We finally would like to comment on how the integrals $I_j$ were
guessed:
   The complicated integral (\ref{a0}) at the end of the
introduction is  quite simple in Moser's coordinates. For example,
$\sqrt{J_1}$ is simply equal to $2(\lambda_2 -\lambda_1)$ and the
expression \bd{ p_1-p_2 + \sqrt{ J_1} \over p_1-p_2-\sqrt{
J_1}}\ed reduces to $-\left( {r_1 \over r_2} \right)^2$. The
exponent is simplified as follows. On the one hand,  \bd {
(q_1+q_2)}^{\dot{}}  =p_1+p_2=-2(b_1+b_2)=-2(\lambda_1+\lambda_2 )
\ . \ed On the other hand from $\dot{r_i} =-r_i \lambda_i$ one
obtains that ${ (\ln r_i)}^{\dot{}}=-\lambda_i$. Therefore, the
function $q_1+q_2$ and $2 \ln r_1 r_2$ differ only by a real
constant. In other words, up to a constant,  the exponent is
simply \bd { 2(\lambda_1-\lambda_2) \over \lambda_1 +\lambda_2}
\ln (r_1 r_2) \ , \ed and   up to a sign difference, the integral
(\ref{a0}) is precisely the same as  the one in (\ref{a9}).
\bigskip

 {\bf Acknowledgements:} The idea to use Moser's
coordinates is due to A. P. Veselov. The authors would like to
thank him for suggesting this approach.

\end{document}